\documentclass[twocolumn,showpacs,superscriptaddress,preprintnumbers,floats,aps,pra,10pt]{revtex4-1}
\usepackage{graphicx}
\usepackage{subfigure}
\usepackage{dcolumn}
\usepackage{amsmath}
\usepackage{amsfonts}
\usepackage{amssymb}
\usepackage{color}
\providecommand{\U}[1]{\protect\rule{.1in}{.1in}}

\begin{document}
\title{Exact analytical solution for an Israel-Stewart Cosmology}
\author{Norman Cruz}
\altaffiliation{norman.cruz@usach.cl}
\affiliation{Departamento de F\'isica, Universidad de Santiago de Chile, \\
Avenida Ecuador 3493, Santiago, Chile.}
\author{Esteban Gonz\'alez}
\altaffiliation{esteban.gonzalezb@usach.cl}
\affiliation{Departamento de F\'isica, Universidad de Santiago de Chile, \\
Avenida Ecuador 3493, Santiago, Chile.}
\author{Guillermo Palma}
\altaffiliation{guillermo.palma@usach.cl}
\affiliation{Departamento de F\'isica, Universidad de Santiago de Chile, \\
Avenida Ecuador 3493, Santiago, Chile.}
\date{\today}

\begin{abstract}
\textbf{{Abstract:}} In this article we report a novel analytic solution for a cosmological model with a matter content described by a one component dissipative fluid, in the framework of the causal Israel-Stewart theory. Some physically well motivated analytical relations for the bulk viscous coefficient, the relaxation time and a bariotropic equation of state are postulated. We study within the parameter space, which label the solution, a suited region compatible with an accelerated expansion of the universe for late times, as well as stability properties of the solution at the critical parameter values $ \gamma = 1$ and for $ s = 1/2 $. We study as well the consequences that arise from the positiveness of the entropy production along the time evolution. In general, the accelerated expansion at late times is only possible when $\epsilon\geq 1/18$, which implies a very large non-adiabatic contribution the speed of sound.

\end{abstract}

\vspace{0.5cm}
\maketitle

\section{Introduction}

As an alternative to $\Lambda CDM$, the DM unified models do not
invoke a cosmological constant. In the framework of general
relativity, non perfect fluids drive accelerated expansion due to
the negativeness of the viscous pressure, which appears from the
presence of bulk viscosity. Therefore, a cold DM viscous component
is a kind of unified DM model that could, in principle, explain the
above mentioned transition without the inclusion of a DE component.

At background level, where a homogeneous and isotropic space
describes the universe as a whole, only bulk viscosity is present in
the cosmic fluid and the dissipative pressure must be described by
some relativistic thermodynamical approach for non perfect fluids.
This implies a crucial point in a fully consistent physical
description of the expansion of the universe using dissipative
processes to generate the transition. Meanwhile, in the $\Lambda
CDM$ model the acceleration is due to a cosmological constant and
the entropy remains constant, in the case of non perfect fluids it
is necessary to find a solution that not only consistently describes
the kinematics of the universe, but also that satisfies the
thermodynamical requirements, such as the positiveness of entropy
generation. In the case of a description of viscous fluids, the
Eckart's theory~\cite{Eckart} has been widely investigated due to
its simplicity and became the starting point to shed some light in
the behavior of the dissipative effects in the late time
cosmology~\cite{Avelino} or in inflationary
scenarios~\cite{Padmanabhan}. In order to avoid superluminal
propagation of the viscous effects and inestabilities, it is
necessary to include a causal description of relativistic non
perfect fluids such as the one given by the Israel-Stewart (IS)
theory~\cite{Israel}-~\cite{Maartens1996}.

We shall assume a barotropic EoS for the one component fluid that filled the universe, with the expression
\begin{equation}
p=\left(\gamma-1\right)\rho,
\end{equation}
where $p$ is the barotropic pressure, $\rho$ is the energy density.
Since our aim is to describe the evolution of the universe with
dissipative normal matter, we shall consider that the EoS parameter
lies in the range $1\leq\gamma <2$.

For the bulk viscous coefficient $\xi(\rho)$ we use the following Ansatz:
\begin{equation}
\xi=\xi_{0}\rho^{s},
\end{equation}
which has been widely considered as a suitable function between the
bulk viscosity and the energy density of the main fluid. $\xi _{0}$
is a positive constant because of the second law of
thermodynamics~\cite{Weinberg1971}.  This particular election of
$\xi$ is rather arbitrary, since we are not considering a
microscopic model of the dark matter that allows, in principle, to
evaluate directly this coefficient from statistical mechanics. On
the other hand, the differential equation for the Hubble parameter
obtained with this Ansatz can be integrated for some particular
values of $s$. In the cases $s= 1/2$ the differential equation is
the most simple to solve.

Taking into account the above assumptions, the IS theory leads to a
nonlinear ordinary differential equation that has been solved and
investigated in many previous works for some particular parameter
values. Using, for example, the factorization method, some new exact
parametric solutions for different values of the viscous parameter
$s$ were found in~\cite{Cornejo}. A particular solution for stiff
matter and $s=1/4$ was found in ~\cite{Harko}. Other exact solutions
found in~\cite{Harko1} well describe determined periods of
inflationary and non inflationary evolutions of the universe.
Inflationary solutions and their stability properties were studied
in~\cite{Chimento}. Using a particular Ansatz for the viscous pressure, 
a solution for the corresponding IS-cosmology is found in ~\cite{Piattella}.

One important assumption in the thermodynamical approaches of
relativistic viscous fluids is the near equilibrium condition, i.e.,
that the viscous pressure must be lower than the equilibrium
pressure of the fluid. In the case of the solutions that present
acceleration from the beginning, like the bulk viscous inflation
case, or at some stage, like those that could represent the late
transition between decelerated and accelerated expansions, the above
condition is not fulfilled, therefore the application of these
theories is not strictly justified. A non linear extension of IS
theory to take into account desviations from the near equilibrium
condition was formulated in~\cite{Maartens1997}. This non linear
extension was investigated in the context of
inflation~\cite{Chimento1} and also in late time phantom
behavior~\cite{Cruzphantom}.

Our novel solution generalizes the exact solution found
in~\cite{Mathew2017}, for the particular values $s=1/2$, $\gamma=1$,
where the expression $\tau=\xi/\rho$ for the relaxation time was
used. In this article, the solution displays a decelerated phase an
exponential expansion for late times, corresponding to a de Sitter
phase. Moreover, our solution was obtained using the following
expression for the relaxation time $\tau$~\cite{Maartens1996},
derived from the study of the causality and stability of the IS
theory in~\cite{Hiscock}
\begin{equation}
\frac{\xi}{\left(\rho+p\right)\tau}=c_{b}^{2},
\label{relaxationtime}
\end{equation}
where $c_{b}$ is the speed of bulk viscous perturbations
(non-adiabatic contribution to the speed of sound in a dissipative
fluid without heat flux or shear viscosity). Since the dissipative
speed of sound $V$, is given by $V^{2}= c_{s}^{2}+c_{b}^{2}$, where
$c_{s}^{2}=(\partial p/\partial \rho)_{s}$ is the adiabatic
contribution, then for a barotropic fluid $c_{s}^{2}=\gamma-1$ and
thus $c_{b}^{2}=\epsilon\left(2-\gamma\right)$ with $0<\epsilon\leq
1$, in order to ensure causality, with a dissipative speed of sound
lower or equal to the speed of light.

In what follows we will discuss our novel solution aiming to obtain
a fully physically consistent behavior within the allowed regions of
the parameters. Our goal is to find a solution in the framework of
unified DM models that can describe consistently well the late
transition between decelerated and accelerated expansion, and, in
addition, presents a behavior consistent with the second law of
thermodynamics, in the context of a linear IS theory. We assume in
the case of accelerated expansion that the linear version of the
causal theory is valid, which occurs within a range of the
parameters involved in our solution.

\section{Israel-Stewart formalism}
In what follows we assume that the universe contains a DM component
which experiments dissipative processes during the cosmic evolution.
We assume a barotropic EoS, $p=\left(\gamma-1\right)\rho$, where $p$
is the barotropic pressure, $\rho$ the energy density and
$1\leq\gamma<2$. For a flat FLRW universe without cosmological
constant, the constraint equation can be written, using  natural
units defined by $8\pi G=c=1$, as
\begin{equation}
3H^{2}=\rho, \label{constraint}
\end{equation}
and the Einstein pressure equation is given by
\begin{equation}
2\dot{H}+3H^{2}=-p-\Pi. \label{acceleration}
\end{equation}
In the IS framework, the transport equation for the viscous pressure $\Pi $ reads~\cite{Israel1979}
\begin{equation}
\tau\dot{\Pi}+\Pi=-3\xi H-\frac{1}{2}\tau\Pi\left(3H+\frac{\dot{\tau}}{\tau }-\frac{\dot{\xi}}{\xi}-\frac{\dot{T}}{T}\right), \label{eqforPi}
\end{equation}
where ``dot" accounts for the derivative with respect to the cosmic
time, $\tau$ is the relaxation time, $\xi(\rho)$ is the bulk
viscosity coefficient, for which we assume a dependence with the
energy density of DM, $H$ is the Hubble parameter and $T$ is the
barotropic temperature, which takes the form
$T=T_{0}\rho^{\left(\gamma-1\right)/\gamma}$ (Gibbs integrability
condition when $p=\left(\gamma-1\right)\rho$) with $T_{0}$ being a
positive parameter. The DM EoS, $\xi(\rho)$ and the relaxation time
are related by Eq.(\ref{relaxationtime}).

It is very interesting and always desirable to obtain analytical
solutions to cosmological models, as they don't suffer from the
numerical instabilities of numerical solutions nor hide a different
underlying behaviour of the dynamical system, implicitly ruled out
by the numerical algorithm used. For this aim we have chosen the
particular case $s=1/2$ and will show a novel exact solution,
discussing its physical properties, in section \textbf{IV}. Thus,
from Eq.(\ref{relaxationtime}) the relaxation time results to be
\begin{equation}
\tau=\frac{\xi_{0}}{\epsilon\gamma\left(2-\gamma\right)}\rho^{s-1}. \label{relaxationtime1}
\end{equation}
In order to obtain a differential equation in terms of the Hubble
parameter, we evaluate the ratios $\dot{\tau}/\tau, \,\,
\dot{\xi}/\xi$ and $\dot{T}/T$, which appear in Eq.(\ref{eqforPi}).
Using Eq.(\ref{constraint}), we get the following expressions
\begin{equation}
\frac{\dot{\tau}}{\tau}=2\left(s-1\right)\frac{\dot{H}}{H}, \label{taupuntotau}
\end{equation}
\begin{equation}
\frac{\dot{\xi}}{\xi}=2s\frac{\dot{H}}{H}, \label{xipuntoxi}
\end{equation}
and
\begin{equation}
\frac{\dot{T}}{T}=2\left(\frac{\gamma-1}{\gamma}\right)\frac{\dot{H}}{H}.  \label{TpuntoT}
\end{equation}
From Eqs.(\ref{constraint}) and (\ref{acceleration}), we obtain the following expression for the viscous pressure
\begin{equation}
\Pi=-\left(2\dot{H}+3\gamma H^{2}\right), \label{Pi}
\end{equation}
whose time derivative is,
\begin{equation}
\dot{\Pi}=-\left(2\ddot{H}+6\gamma H\dot{H}\right). \label{Pipunto}
\end{equation}
Finally, inserting Eqs.(\ref{relaxationtime1}-\ref{Pipunto}) into
Eq.(\ref{eqforPi}), we obtain the nonlinear second order
differential equation for $H$, that represents the general
differential equation to be solved in this model, which governs the
time evolution of the Hubble parameter
\begin{widetext}
\begin{equation}
\begin{split}
& \ddot{H}+3H\dot{H}+(3)^{1-s}\xi_{0}^{-1}\epsilon\gamma\left(2-\gamma\right)H^{2-2s}\dot{H}-\frac{(2\gamma-1)}{\gamma}H^{-1}\dot{H}^{2}+\frac{9}{4}\gamma\left[1-2\epsilon\left(2-\gamma\right)\right]H^{3} \\
& +\frac{1}{2}(3)^{2-s}\xi_{0}^{-1}\epsilon\gamma^{2}\left(2-\gamma\right)H^{4-2s}=0. \label{eqforH}
\end{split}
\end{equation}
\end{widetext}

In the special case where $s=1/2$, Eq.(\ref{eqforH}) has a phantom
solution of the form $H\left(t\right)=A\left(t_{s}-t\right)^{-1}$,
with $A>0$, $\epsilon=1$ and the restriction $ 0<\gamma<3/2$. This
solution was discussed in~\cite{Cruz2017}. Also the solution
$H\left(t\right)=A\left(t-t_{s}\right)^{-1}$ can represent
accelerated universes if $A>1$, Milne universes if $A=1$ and
decelerated universes if $A<1$, all with an initial singularity at
$t=t_{s}$~\cite{Cruz2017a}. It is worthy mentioning that only the
decelerated solution satisfies a positive entropy production,
therefore there is no transition from a decelerated phase to an
accelerated one, as it occurs in the standard model. As we shall see
below, the dynamical behavior of an exact solution of a model
described by the IS thermodynamic formalism does not necessarily
implies that its thermodynamical properties behave physically
consistent.

\section{De Sitter type like solution}
There is a mathematically trivial solution of Eq.(\ref{eqforH}) for
the special value $s=1/2$, which is known as a de Sitter type
solution, which coincides with the asymptotic behavior of the
$\Lambda CDM$ model. In fact, for $H=\textup{const}$ a solution of
Eq.(\ref{eqforH}) reads:
\begin{equation}
H=\left\{\frac{3^{s}\xi_{0}}{2}\left[\frac{2\epsilon\left(2-\gamma\right)-1}{\epsilon\gamma\left(2-\gamma\right)}\right]\right\}^{\frac{1}{1-2s}}. \label{HdeSitter}
\end{equation}

It is easy to see that there is no de Sitter solution when $s=1/2$
as the exponent flows up. On the other hand, if we require a
positive Hubble parameter that represents an expanding universe (or
avoids a complex Hubble parameter) we need to impose that the term
within parenthesis be positive. Because $\epsilon=0$ and $\gamma=2$
indeterminate the Hubble parameter, we have to restrict the
parameters to the regions $0<\epsilon\leq 1$ and $1\leq\gamma<2$.
Furthermore as $\xi_{0}>0$, an expanding universe requires
\begin{equation}
\frac{1}{2}\leq\frac{1}{2\left(2-\gamma\right)}<\epsilon\leq 1 \;\; \textup{with} \;\; 1\leq\gamma<\frac{3}{2}. \label{constraintdeSitter}
\end{equation}

The solution of Eq.(\ref{HdeSitter}) was previously found
in~\cite{Cruz2017}, but the particular value $\epsilon=1$ was used,
so the lower bound for $\epsilon$ displayed in
(\ref{constraintdeSitter}) was missing.

\section{A novel analytical solution for arbitrary $\gamma$}
A new analytical solution can be found for the Eq.(\ref{eqforH}) if
we consider the particular value $s=1/2$. In fact in this case
Eq.(\ref{eqforH}) goes into
\begin{equation}
\ddot{H}+d_{1}H\dot{H}+d_{2}H^{3}-d_{3}\frac{\dot{H}^{2}}{H}=0, \label{difeqforHtgamma}
\end{equation}
where for simplicity we have defined the constants
\begin{equation}
d_{1}\equiv 3\left[1+\frac{\epsilon\gamma\left(2-\gamma\right)}{\sqrt{3}\xi_{0}}\right], \label{defofd1}
\end{equation}
\begin{equation}
d_{2}\equiv \frac{9}{4}\gamma\left\{\left[1-2\epsilon\left(2-\gamma\right)\right]+\frac{2\epsilon\gamma\left(2-\gamma\right)}{\sqrt{3}\xi_{0}}\right\}, \label{defofd2}
\end{equation}
\begin{equation}
d_{3}\equiv\frac{2\gamma-1}{\gamma}. \label{defofd3}
\end{equation}

In the Eq.(\ref{difeqforHtgamma}) we change the variable from the cosmic time $t$ to $x=\ln{\left(a\right)}$, and the differential
equation takes the form
\begin{equation}
\dfrac{d^{2}H}{dx^{2}}+d_{1}\dfrac{dH}{dx}+d_{2}H+\frac{1-d_{3}}{H}\left(\frac{dH}{dx}\right)^{2}=0, \label{difeqforHxgamma}
\end{equation}
which is a nonlinear second order differential equation. Further using the Ansatz
\begin{equation}
H(x)=e^{-\frac{d_{1}x}{2\left(2-d_{3}\right)}}\phi\left(x\right), \label{ansatz1}
\end{equation}
Eq.(\ref{difeqforHxgamma}) goes into the equation
\begin{equation}
\dfrac{d^{2}\phi}{dx^{2}}+\left[d_{2}-\frac{d_{1}^{2}}{4\left(2-d_{3}\right)}\right]\phi+\frac{\left(1-d_{3}\right)}{\phi}\left(\dfrac{d\phi}{dx}\right)^{2}=0, \label{difeqforansatz1}
\end{equation}
i.e. we have eliminated the linear first derivative term. Now, in
order to eliminate the nonlinear term in the above equation, we use
a nonlinear second Ansatz
\begin{equation}
\phi\left(x\right)=\left[\Phi\left(x\right)\right]^{\frac{1}{2-d_{3}}}, \label{ansatz2}
\end{equation}
and Eq.(\ref{difeqforansatz1}) reduces to the following expression
\begin{equation}
\dfrac{d^{2}\Phi}{dx^{2}}-\frac{1}{4}\left[d_{1}^{2}+4d_{2}\left(d_{3}-2\right)\right]\Phi=0, \label{difeqforansatz2}
\end{equation}
which is in fact a linear second order differential equation. Thus,
the general solution of Eq.(\ref{difeqforHxgamma}) can be expressed
as
\begin{equation}
H\left(x\right)=e^{-\alpha x}\left[A\cosh{\left(\beta x\right)}+B\sinh{\left(\beta x\right)}\right]^{\gamma}, \label{solforHgamma}
\end{equation}
where $A$, $B$ are integration constants, and

\begin{equation}
\alpha=\frac{\sqrt{3}\gamma}{2\xi_{0}}\left[\sqrt{3}\xi_{0}+\frac{\epsilon\gamma}{3}\left(2-\gamma\right)\right]
\label{defofalpha}
\end{equation}
\begin{equation}
\beta=\frac{\sqrt{3}}{2\xi_{0}}\sqrt{6\xi_{0}^{2}\epsilon\left(2-\gamma\right)+\epsilon^{2}\gamma^{2}\left(2-\gamma\right)^{2}}. \label{defofbeta}
\end{equation}

\subsection{Mathematical properties of the solution and the Liapunov stability of the $\gamma = 1$-limit}
Before studying the behaviour of the Hubble parameter obtained above, it is worthwhile discussing some interesting mathematical properties of the solutions.

Note that the Eq.(\ref{difeqforHxgamma}) is scale-invariant, i. e., if we perform the conformal change $H(x)\rightarrow \sigma H(x)$ for $\sigma$ constant, then the differential equation remains unchanged. We therefore look for a solution of the form
\begin{equation}
H\left(x\right)=e^{\lambda x}, \label{solforinvariant}
\end{equation}
which leads to the following condition on the constant $\lambda$
\begin{equation}
\lambda_{\pm}=\frac{-d_{1}\pm\sqrt{d_{1}^{2}+4d_{2}\left(d_{3}-2\right)}}{2\left(2-d_{3}\right)}. \label{conditionlambda}
\end{equation}
Because Eq.(\ref{difeqforHxgamma}) is a non linear differential equation, then the superposition principle does not hold. Nevertheless, from Eqs.(\ref{solforinvariant}) and (\ref{conditionlambda}), there are two (linearly) independent solutions
\begin{equation}
H_{+}\left(x\right)=e^{\lambda_{+}x} \;\;\; \textup{and} \;\;\; H_{-}\left(x\right)=e^{\lambda_{-}x}, \label{solforlambda}
\end{equation}
but as already mentioned, a linear combination of them does not in general fulfil the differential equation.

In order to find a general solution of the second order differential equation, we need to explore the conditions under which a general linear combination of the solutions (\ref{solforlambda}) is also a solution. To this aim we consider
\begin{equation}
H\left(x\right)=C_{1}H_{+}\left(x\right)+C_{2}H_{-}\left(x\right), \label{Hlinear}
\end{equation}
and inserting this into Eq.(\ref{difeqforHxgamma}) we obtain the following condition on the parameters defined in Eqs.(\ref{defofd1})-(\ref{defofd3})
\begin{equation}
\left(\lambda_{+}^{2}+\lambda_{-}^{2}\right)+d_{1}\left(\lambda_{+}+\lambda_{-}\right)+2d_{2}+2\left(1-d_{2}\right)\lambda_{+}\lambda_{-}=0. \label{conditionalmbda2}
\end{equation}
This condition does not imply a constraint on the constant $C_{1}$ and $C_{2}$, but leads to a new condition on the free parameters $\epsilon$, $\gamma$ and $\xi_{0}$. After some computations, the condition of (\ref{conditionalmbda2}) can be written as
\begin{equation}
\frac{\left(1-d_{3}\right)}{\left(2-d_{3}\right)}\left[4d_{2}-\frac{d_{1}^{2}}{\left(2-d_{3}\right)}\right]=0. \label{linearcondition}
\end{equation}
From the above equation there are two possibilities. The first on is,
\begin{equation}
\frac{c_{b}^{2}\gamma^{2}}{6\xi_{0}^{2}}=-1, \label{linaerrestriction1}
\end{equation}
which clearly cannot be fulfilled for real parameters. The second possibility leads to the condition
\begin{equation}
1-d_{3}=0, \label{linearrestrcition2}
\end{equation}
which implies $\gamma=1$. Thus, the linear combination is a solution
of Eq.(\ref{difeqforHxgamma}) only when $\gamma$ has the particular
value $1$. But for this particular value the nonlinear term of
Eq.(\ref{difeqforHxgamma}) vanishes and leads to a second order
linear differential equation, whose solutions are indeed
exponentials, and are trivially given by the linear combination of
the form given by Eq.(\ref{Hlinear}), but we the modified values of
the parameters $\bar{\lambda}_{\pm}$ given by

\begin{equation}
\bar{\lambda}_{\pm}=\frac{-\bar{d_{1}}\pm\sqrt{\bar{d_{1}}^{2}-4\bar{d_{2}}}}{2}, \label{lambda_pm}
\end{equation}
where, $\bar{d_1} = 3( 1+ \epsilon / \sqrt{3}\xi_{0} )$ and $\bar{d_2} = 9/4(1-2\epsilon + 2\epsilon /\sqrt{3}\xi_{0})$.

\vspace{0.3cm}

In the $\gamma=1$-limit Eq.(\ref{difeqforHxgamma}) has the
remarkable property that the trivial solution $ H(x)=0$ is
asymptotically stable or Liapunov stable, as can be seen by
rewriting it as the differential first order system

\begin{equation}
\dfrac{d u}{d x} = v, \label{syst_first}
\end{equation}
\begin{equation}
\dfrac{d v}{d x} = -\bar{d_{2}}u -\bar{d_{1}}v, \label{syst_second}
\end{equation}
where we have defined $u(x)=H(x)$ and $v(x) = du/dx$. The roots of
the characteristic secular equation associated to this system are
precisely $ \bar{\lambda}_{\pm}$ defined above, and as $ \bar{d_{1}}
> 0$ and $ \bar{d_{1}}^{2}-4\bar{d_{2}} = 9(2\epsilon
+{(\epsilon/\xi_{0})}^{2}/3)< \bar{d_{1}}$, we conclude that both
eigenvalues are real and positive. This is equivalent to the
Liapunov stability of the system, or from the physical point of
view, the solutions are stable under small changes (uncertainty) in
the initial values $H(0)$ and $\dot{H}(0)$. For completeness sake,
we write explicitly the solutions of the system

\begin{equation}
u(x) = a_1 \alpha_{1} \exp(\bar{\lambda}_{+}) + a_2 \beta_{1} \exp(\bar{\lambda}_{-}), \label{u_homogen}
\end{equation}
\begin{equation}
v(x) = a_1 \alpha_{2} \exp(\bar{\lambda}_{+}) + a_2 \beta_{2} \exp(\bar{\lambda}_{-}), \label{v_homogen}
\end{equation}
where $a_1$ and $a_2$ are arbitrary constants, $\alpha_{1}=1= \beta_{1}$, $\alpha_{2}=\bar{\lambda}_{+}$, and finally $\beta_{2}=\bar{\lambda}_{-}$.

Now we want to study whether this property is preserved or not by
the nonlinear term of Eq.(\ref{difeqforHxgamma}). In order to
address this issue, we will consider a perturbative analysis in a
vicinity of $\gamma=1$ by setting $\gamma = 1+ \delta$, with $\delta
\ll 1$.  We further use the following Ansatz for the Hubble
parameter
\begin{equation}
H_{\delta}\left(x\right)=e^{\bar{\lambda}_{\pm} x}\left[\hspace{0.05 cm} 1+ \delta \hspace{0.05 cm} \omega\left(x\right)\right]. \label{H_perturbative}
\end{equation}

Inserting the above Ansatz into Eq.(\ref{difeqforHxgamma}) one obtains the perturbative first order equation  for $\omega$:
\begin{equation}
\dfrac{d^{2}\omega}{dx^{2}}+(2 \bar{\lambda}_{\pm}+\bar{d_{1}})\dfrac{d\omega}{dx}+\alpha=\bar{\lambda}_{\pm}^2 \exp(\bar{\lambda}_{\pm} x). \label{first_order}
\end{equation}
After the integration of the above equation, using for instance the
Cauchy's formula, one finds up to irrelevant additive constants

\begin{equation}
\omega(x) =  \frac{1}{\bar{\lambda}_{\pm} \pm (\bar{d_{1}}^{2}-4\bar{d_{2}})^{1/2}} \hspace{0.1cm} \exp(\bar{\lambda}_{\pm} x).  \label{first_orde_omegar}
\end{equation}

It is worthwhile pointing out that $\omega (x)$ has the same form as
the unperturbated solution (\ref{solforinvariant}) for $\lambda $
replaced by $\bar{\lambda}$. Inserting the analytic expression for
$\omega(x)$ into Eq.(\ref{H_perturbative}) one sees that the
exponent of the Hubble's parameter remains negative, which leads to
the conclusion that the associated system is exponentially stable or
Liapunov stable up to first order in $\delta$.

Moreover, up to first order the nonlinear term in
Eq.(\ref{difeqforHxgamma}) does not change the behavior of the
$\gamma=1$-solution and therefore the scale-invariant solution
(\ref{solforinvariant}) is perturbatively stable in the sense that
\begin{equation}
\lim_{x\rightarrow\infty}{\frac{\left(H_{\delta}-H_{0}\right)\left(x\right)}{H_{0}\left(x\right)}}=0. \label{stablesense}
\end{equation}
In other words, the nonlinear contribution of
Eq.(\ref{difeqforHxgamma}) does not change the asymptotic behavior
of $H\left(x\right)$ up to first order in $\delta$.

\subsection{Behavior of the scale factor}
In what follows we find an implicit solution for the scale factor
$a(t)$. From the definition $H=\dot{a}/a$, Eq.(\ref{solforHgamma})
leads to the implicit integral representation
\begin{equation}
t+C=\frac{1}{C_{3}}\int\frac{a^{1-\alpha} \hspace{0.2cm} da}{\left(D a^{\beta}/2 + E a^{-\beta}/2\right)^{\gamma}},
\label{intforagamma}
\end{equation}
where $C_{3}$ is another integration constant. The above integral
can be expressed as an hyper-geometric function $_{2}F_{1}$, in
particular considering the initial condition $a(t_{0})=1$, the scale
factor is given by the following implicit expression
\begin{equation}
\begin{split}
& a^{\alpha+\gamma\beta}\,_{2}F_{1}\left[\gamma,\frac{\alpha+\gamma\beta}{2\beta},1+\frac{\alpha+\gamma\beta}{2\beta},- \frac{D}{E} a^{2\beta}\right]= \\
& _{2}F_{1}\left[\gamma,\frac{\alpha+\gamma\beta}{2\beta},1+\frac{\alpha+\gamma\beta}{2\beta},-\frac{D}{E}\right] \\
& +\frac{C_{3} \left(\alpha+\gamma\beta\right)}{2^{\gamma}E^{-\gamma}}\left(t-t_{0}\right).
\end{split} \label{solforagamma}
\end{equation}
Due to the complexity of numerically solving the above equation, the
expansion behaviour of the universe will be done instead by
considering the dynamical evolution of the Hubble parameter $H$, and
the deceleration parameter $q = -(1+\dot{H}/H^2)$. Using the
expression for $H(a)$ given by Eq.(\ref{solforHgamma}), $q$ can be
expressed as

\begin{equation}
q = -1-\gamma \frac{\left( \frac{D}{2} \lambda_{+} \hspace{0.1cm} a^{\lambda_{+}} + \frac{E }{2} \lambda_{-}\hspace{0.1cm} a^{\lambda_{-}}\right)}{\left(\frac{D}{2} \hspace{0.1cm} a^{\beta} + \frac{E}{2} \hspace{0.1cm} a^{-\beta}\right)},
\label{qofzgamma1}
\end{equation}
where $ \lambda_{\pm}$ was defined in Eq.(\ref{conditionlambda}).

Now, in order to simplify the above expressions, we make the
particular choice of parameters $D = \exp(\beta C_{4})$, $E =
\exp(-\beta C_{4})$ and use the redshift $z$ defined as usual by $z
= 1/a-1$, in which the limit $z\rightarrow\infty$ corresponds to
very early times while $z\rightarrow -1$ represents the very far
future. With these choices, the Hubble and deceleration parameters
have respectively the following compact forms

\begin{equation}
H(z) = C_{3} \hspace{0.1cm} \left(1+z\right)^{\alpha}
\cosh^{\gamma}
{\left[\beta\left(\ln{\left(1+z\right)}+C_{4}\right)\right]},
\label{Hzgamma}
\end{equation}
\begin{equation}
q(z)=-1+\alpha+\gamma\beta\tanh{\left[\beta\left(\ln{\left(1+z\right)}+C_{4}\right)\right]}, \label{qofzgamma}
\end{equation}
where $C_{3}$ and $C_{4}$ are constants given by
\begin{equation}
C_{3}=\frac{H_{0}}{\cosh^{\gamma}{\left(\beta C_{4}\right)}}=H_{0}\left[1-\frac{\left(q_{0}+1-\alpha\right)^{2}}{\gamma^{2}\beta^{2}}\right]^{\frac{\gamma}{2}}, \label{defofC3}
\end{equation}
\begin{equation}
C_{4}=\frac{1}{\beta}\mathop{\mathrm{arctanh}}\left[\frac{\left(q_{0}+1\right)-\alpha}{\gamma\beta}\right]. \label{defofC4}
\end{equation}
In the above equations $H_{0}$ and $q_{0}$ are the Hubble and the deceleration parameters respectively, at the present time $t=t_{0}$. We have also set the condition $a_{0}=1$.

Note from the Eqs.(\ref{defofC3}) and (\ref{defofC4}) that for a real Hubble parameter the deceleration parameter $q_{0}$ must fulfill the constraints
\begin{equation}
\left(\alpha-\gamma\beta\right)-1<q_{0}<\left(\alpha+\gamma\beta\right)-1. \label{constraintq0gamma}
\end{equation}
A consequence of the above restriction is that the Hubble parameter given by the Eq.(\ref{Hzgamma}) remains positive during the whole cosmic evolution. In Fig. 1 are displayed the allowed regions imposed by the constraint in terms of the parameters $q_{0}$, $\xi_{0}$, $\gamma$ and $\epsilon$.

The asymptotic behaviour can be easily computed with the above expressions. For early times it holds
\begin{equation}
H\left(z\rightarrow\infty\right)\rightarrow C_{3}\left(\frac{e^{\beta C_{4}}}{2}\right)^{\gamma}\left(1+z\right)^{\alpha+\gamma\beta}, \label{Hgammaearly}
\end{equation}
\begin{equation}
q\left(z\rightarrow\infty\right)\rightarrow -1+\left(\alpha+\gamma\beta\right), \label{qofzgammaearly}
\end{equation}
while for the very far future they behave as
\begin{equation}
H\left(z\rightarrow -1\right)\rightarrow C_{3}\left(\frac{e^{-\beta C_{4}}}{2}\right)^{\gamma}\left(1+z\right)^{\alpha-\gamma\beta}, \label{Hgammlate}
\end{equation}
\begin{equation}
q\left(z\rightarrow -1\right)\rightarrow -1+\left(\alpha-\gamma\beta\right). \label{qofzgammalate}
\end{equation}

Note that the behaviour of the Eqs.(\ref{Hgammaearly}) and (\ref{qofzgammaearly}) depends on the exponent $\left(\alpha+\gamma\beta\right)$ defined by Eqs.(\ref{defofalpha}) and (\ref{defofbeta}), which is always positive. Furthermore, this exponent has the following constraint
\begin{equation}
\alpha+\gamma\beta\geq\frac{3\gamma}{2}. \label{exponentearlyconstraint}
\end{equation}
Therefore, the Hubble parameter is positive at early times, and monotonically decreasing with the redshift. This behavior corresponds to a decelerated expansion, as it can be see from Eqs.(\ref{qofzgammaearly}) and (\ref{exponentearlyconstraint}), which leads to a lower bound for the deceleration parameter $q\geq 1/2$.

On the other hand, the behavior of the Eqs.(\ref{Hgammlate}) and (\ref{qofzgammalate}) is driven by the exponent $\left(\alpha-\gamma\beta\right)$, which is positive for
\begin{equation}
3\left[1-2\epsilon\left(2-\gamma\right)\right]\xi_{0}+2\sqrt{3}\epsilon\gamma\left(2-\gamma\right)>0,
\label{conditionpowerpositive}
\end{equation}
where the expressions for $\alpha$ and $\beta$ are given by Eqs. (\ref{defofalpha}) and (\ref{defofbeta}) respectively, in terms of the parameters $\gamma$, $\epsilon$ and $\xi_{0}$. It follows that for $\gamma\geq 3/2$,  $\alpha-\gamma\beta$ it is positive, and from Eq.(\ref{Hgammlate}) we conclude that the Hubble parameter goes to zero in the infinite cosmological time limit. The same behavior arises when $1\leq\gamma<3/2$, with $0<\epsilon\leq\frac{1}{2\left(2-\gamma\right)}$ and for $\frac{1}{2\left(2-\gamma\right)}<\epsilon\leq 1$, if and only if, $\xi_{0}$ satisfies the additional inequality
\begin{equation}
\xi_{0}<\frac{2\epsilon\gamma\left(2-\gamma\right)}{\sqrt{3}\left[2\epsilon\left(2-\gamma\right)-1\right]}. \label{constraintxi0powerpositive}
\end{equation}
On the other hand, if $1\leq\gamma<3/2$, then $\alpha-\gamma\beta<0$ for $0<\epsilon\leq\frac{1}{2\left(2-\gamma\right)}$, and if the constraint (\ref{constraintxi0powerpositive}) is not satisfied, the Hubble parameter at late times stop to decrease and start to grow, becoming infinite at $z=-1$. Finally, if $1\leq\gamma<3/2$, then we have the especial case in what $\alpha-\gamma\beta=0$ when $0<\epsilon\leq\frac{1}{2\left(2-\gamma\right)}$ and if the inequality in the Eq.(\ref{constraintxi0powerpositive}) becomes an equality, and we will have a constant Hubble parameter at late times.

\begin{figure}[ht]
\centering
\subfigure[\hspace{0.1cm} Plot of the deceleration parameter as a function of the redshift for a fixed $\gamma=1.1$ value. The full line correspons to $\epsilon=0.8$, $\xi_{0}=1.05$ and $q_{0}=-0.6$; the dashed line corresponds to $\epsilon=0.4$, $\xi_{0}=0.8$ and $q_{0}=0.1$; and dashed-dotted line one corresponds to the values $\epsilon=0.1$, $\xi_{0}=0.2$ and $q_{0}=0.5$.]{\includegraphics[scale=0.68]{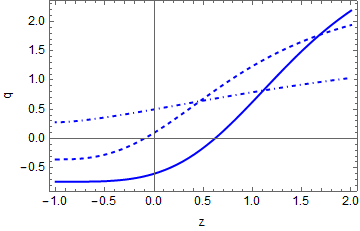}}
\vspace{0.05cm}
\subfigure[\hspace{0.1cm}Contour plot of the allowed values for the free parameters $\xi_{0}$ and $\epsilon$ for fixed $\gamma$ that leads to a transition between the decelerated expansion to a accelerated one, which occurs at $z=0.64$ with $q_{0}=-0.6$. The full line is for $\gamma=1.01$, dashed line is for $\gamma=1.05$ and dashed-doted one corresponds to $\gamma=1.1$.]{\includegraphics[scale=0.68]{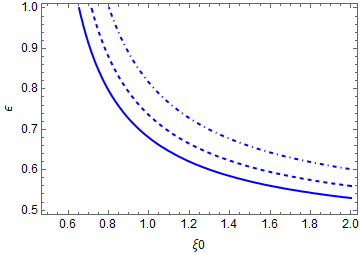}}
\caption{Plot of the deceleration parameter as a function of the redshift for a fixed $\gamma$ value (a), and contour plot of the allowed values for the free parameters that leads to a transition between the decelerated expansion to a accelerated one for three different $\gamma$ values (b).}
\label{fig:q0behavior}
\end{figure}

From Eq.(\ref{qofzgammalate}) we pointed out that, this last behavior leads to a accelerated expansion, provided the following two conditions are fulfilled
\begin{equation}
\begin{split}
\xi_{0}> & \epsilon\gamma\left(2-\gamma\right)\left(\frac{6\gamma-4}{\sqrt{3}\gamma}\right)\times \\
& \left[\frac{3\gamma^{2}}{18\epsilon\gamma^{2}\left(2-\gamma\right)-9\gamma^{2}+12\gamma-4}\right], \label{constraintxi0acceleratedgamma}
\end{split}
\end{equation}
and
\begin{equation}
\epsilon>\frac{9\gamma^{2}-12\gamma+4}{18\gamma^{2}\left(2-\gamma\right)}. \label{constraintepsilonacceleratedgamma}
\end{equation}
If one of the above conditions is not fulfilled, then the behavior of the Hubble parameter leads to a decelerated expansion. The transition between the accelerated expansion to a decelerated one occurs at redshift value
\begin{equation}
z_{q=0}=-1+exp\left[\frac{1}{\beta}\mathop{\mathrm{arctanh}}\left(\frac{1-\alpha}{\gamma\beta}\right)-C_{4}\right]. \label{zqequalcerogamma}
\end{equation}
It is important to mention that in all cases the constraint of Eq.(\ref{constraintq0gamma}) must be fulfilled. In Fig.\ref{fig:q0behavior}a) the behavior of the deceleration parameter is displayed in terms of the free parameters $q_{0}$, $\xi_{0}$, $\gamma$ and $\epsilon$. Note that only for large values of epsilon and a negative $q_{0}$ is possible to obtain a transition in the past and for a z value compatible with the observations. In Fig.\ref{fig:q0behavior}b) we have use Eq.(\ref{zqequalcerogamma}) to draw the allowed values for $\xi_{0}$ and $\epsilon$, for fixed gamma, where we have chosen the evaluated value value from observations for the transition redshift, $z=0.64$, and also the estimated value for $q_{0}$ at the present time: $q_{0}=-0.6$. It can be also noted that the transition occurs only for large values of $\epsilon$, or, instead of this, for very large values of $\xi_{0}$.

\subsection{Thermodynamical properties of the solution}
In this section we will evaluate the entropy production due to the dissipative process. For this aim the following relation is used
\begin{equation}
\dfrac{dS}{dt}=-\frac{3H\Pi}{nT}, \label{entropyproduction}
\end{equation}
where $n$ is the number of particles, which has to satisfy the conservation equation
\begin{equation}
\dot{n}+3Hn=0. \label{conservationeq}
\end{equation}
The solution of the above equation in terms of the scale factor is
\begin{equation}
n=\frac{n_{0}}{a^{3}}. \label{particlenumber}
\end{equation}
Using Eq.(\ref{constraint}), one sees that $T=T_{0}\left(3H^{2}\right)^{\left(\gamma-1\right)/\gamma}$, and from Eqs.(\ref{Pi}) and (\ref{particlenumber}) we can rewrite Eq.(\ref{entropyproduction}) in the form
\begin{equation}
\dfrac{dS}{dt}=\frac{3^{1/\gamma}a^{3}}{T_{0}n_{0}}H^{2/\gamma}\left(2\dfrac{dH}{da}a+3\gamma
H\right). \label{entropyproduction1}
\end{equation}
Now using the expression of Eq.(\ref{Hzgamma}) for the Hubble parameter, the above equation can be finally written as
\begin{equation}
\begin{split}
& \dfrac{dS}{dt}=\frac{3^{1/\gamma}a^{3}}{T_{0}n_{0}}H^{\left(2+\gamma\right)/\gamma}\times \\
&
\left[-2\gamma\beta\tanh{\left[\beta\left(\ln{\left(1+z\right)+C_{4}}\right)\right]-2\alpha+2\gamma}\right].
\label{entropyofagamma}
\end{split}
\end{equation}
Because of the second law of thermodynamics, the entropy production must be a non-negative function of the time. This requirement constraints the parameters of the r.h.s. of the above equation, which leads to the condition
\begin{equation}
-2\gamma\beta\tanh{\left[\beta\left(\ln{\left(1+z\right)+C_{4}}\right)\right]-2\alpha+3\gamma}\geq 0. \label{entropygamma}
\end{equation}

Similarly to what was already done for the Hubble parameter, we will analyze the above condition only for early and very far future times. This is why we will only consider the strict inequality of Eq.(\ref{entropygamma}), and we will study its saturation $ds/dt=0$ only if it is required. It is easy to note that if $z\rightarrow\infty$ then the term inside the brackets tends to the constant expression
\begin{equation}
\begin{split}
& 3\gamma-2\left(\alpha+\gamma\beta\right)=-\frac{\sqrt{3}\gamma}{\xi_{0}}\times \\
& \left[\epsilon\gamma\left(2-\gamma\right)+\sqrt{6\xi_{0}^{2}\epsilon\left(2-\gamma\right)+\epsilon^{2}\gamma^{2}\left(2-\gamma\right)^{2}}\right]<0. \label{asymptoticsgamma1}
\end{split}
\end{equation}
On the other hand, if $z\rightarrow -1$, the term within the brackets tends to the constant expression
\begin{equation}
\begin{split}
& 3\gamma+2\left(\gamma\beta-\alpha\right)=\frac{\sqrt{3}\gamma}{\xi_{0}}\times \\
& \left[\sqrt{6\xi_{0}^{2}\epsilon\left(2-\gamma\right)+\epsilon^{2}\gamma^{2}\left(2-\gamma \right)^{2}}-\epsilon\gamma\left(2-\gamma\right)\right]>0.
\end{split}. \label{asymptoticsgamma2}
\end{equation}
Thus, Eqs.(\ref{asymptoticsgamma1}) and (\ref{asymptoticsgamma2}) show that the entropy production is negative at early times and positive for late times, which leads to the conclusion that this model is not fully consistent with the physical requirement of an entropy monotonically growing in the whole range of the cosmological time. Nevertheless, this solution has been considered from the very beginning with only one matter fluid, which we expect to successfully describes the transition from decelerated to accelerated expansion, but as it does not include the contribution from radiation, that is necessary to consider in order to describe early times of the universe. From Eq.(\ref{entropygamma}) it follows that the change of sign in the entropy production occurs at a redshift value given by
\begin{equation}
z_{ds/dt=0}=-1+exp\left[\frac{1}{\beta}\mathop{\mathrm{arctanh}}\left(\frac{2\alpha-3\gamma}{2\gamma\beta}\right)-C_{4}\right]. \label{zentropyequalcerogamma}
\end{equation}
Therefore, our solution at late times can successfully describe, for certain particular parameters values, the above mentioned transition and furthermore has a positive entropy production. Of course, at late times the dominant fluid is the pressureless DM, and therefore have to analyze the particular solution with $\gamma =1$, which is addressed in the following section. A numerical calculation of Eq.(\ref{zentropyequalcerogamma}) indicates us that the the transition from a negative entropy production to a positive one may occurs at z in the range $1<z<5$ choosing values of epsilon between $0.5$ and $0.7$, and $0.8<\xi_{0}<2$. In other words,  allowed values of the model's parameters  can describe an scenario where the transition from a decelerated expansion to an accelerated one, occurs while the entropy production remains positive.

\section{The particular case $\gamma =1$}
As it was observed in the section \textbf{IV.A.}, when $\gamma=1$ or, in other words, when a pressureless DM is considered as the main material content of the universe, a particular solution of Eq.(\ref{difeqforHtgamma}) is obtained by Eq.(\ref{Hlinear}). Considering Eqs.(\ref{solforinvariant}) and (\ref{conditionlambda}) with $\gamma=1$, and recalling that $x=\ln{\left(a\right)}=-\ln{\left(1+z\right)}$, this solution can be written as
\begin{equation}
H(z)=H_{0}\left[C_{1}\left(1+z\right)^{m_{1}}+C_{2}\left(1+z\right)^{m_{2}}\right], \label{solforH}
\end{equation}
where $H_{0}$ is the Hubble parameter at the present time $t=t_{0}$, and
\begin{equation}
m_{1}=\frac{\sqrt{3}}{2\xi_{0}}\left(\sqrt{3}\xi_{0}+\epsilon+\sqrt{6\xi_{0}^{2}\epsilon+\epsilon^{2}}\right), \label{defofm1}
\end{equation}
\begin{equation}
m_{2}=\frac{\sqrt{3}}{2\xi_{0}}\left(\sqrt{3}\xi_{0}+\epsilon-\sqrt{6\xi_{0}^{2}\epsilon+\epsilon^{2}}\right), \label{defofm2}
\end{equation}
\begin{equation}
C_{1}=\frac{\left(q_{0}+1\right)-m_{2}}{m_{1}-m_{2}}, \label{defofc1}
\end{equation}
\begin{equation}
C_{2}=\frac{m_{1}-\left(q_{0}+1\right)}{m_{1}-m_{2}}. \label{defofc2}
\end{equation}
In the above equations $q_{0}$ is the deceleration parameter at the present time $t=t_{0}$, and the conditions $a_{0}=1$ and $C_{1}+C_{2}=1$ have been set. This solution was previously found and discussed in~\cite{Mathew2017}, but with a particular relation for the relaxation time of the form $\xi_{0}\rho^{s-1}$ (which correspond to $\alpha=\xi_{0}$ for our), instead of the more general relation as Eq.(\ref{relaxationtime1}), in which the causality condition $0<\epsilon\leq 1$ is imposed. From a perturbative point of view, it is necessary to have a knowledge of the speed of sound in the fluid, which has to be very close to zero in order to be compatible with the growth of structures.  In this sense, imposing
from the beginning $\epsilon=1$ leads to possible solutions of the Israel-Stewart equation that could behave reasonable at the background level, but present drawbacks at perturbative level.

Furthermore, in the solution found in~\cite{Mathew2017}, $\Pi$ was used as a second initial condition, instead of using $q_{0}$. The constants $m_{1}$ and $m_{2}$ are reduced to the corresponding constants in~\cite{Mathew2017} by taking the particular value $\epsilon=1$. On the other hand, the Hubble parameter given by the Eq.(\ref{Hzgamma}) goes into the expression for the Hubble parameter given by Eq.(\ref{solforH}) when one chooses $\gamma=1$. It is important to mention that in this solution there is no restriction upon $q_{0}$, being the most important feature of the new analytical solution.

\subsection{Dynamics of the universe}
In this subsection we are interested in characterizing the expansion of the universe according to the particular solution found in section \textbf{V} for the particular value $\gamma =1$, which corresponds to CDM. From the definition $H=\dot{a}/a$ and from Eq.(\ref{solforH}) it follows that
\begin{equation}
H_{0}t+C=\int{\frac{a^{m_{1}+m_{2}-1}}{C_{1}a^{m_{2}}+C_{2}a^{m_{1}}}da}, \label{intfora}
\end{equation}
where $C$ is an integration constant. Similarly as the method used in the section \textbf{IV-B}, the integral of the above equation can be expressed as a hyper-geometric function $_{2}F_{1}$. For the initial condition $a(t_{0})=1$, the scale factor is given by the following implicit formula
\begin{equation}
\begin{split}
& a^{m_{1}}\;_{2}F_{1}\left[1,\frac{m_{1}}{m_{1}-m_{2}},1+\frac{m_{1}}{m_{1}-m_{2}},-a^{m_{1}-m_{2}}\frac{C_{2}}{C_{1}}\right]= \\
& _{2}F_{1}\left[1,\frac{m_{1}}{m_{1}-m_{2}},1+\frac{m_{1}}{m_{1}-m_{2}},-\frac{C_{2}}{C_{1}}\right] \\
& +C_{1}m_{1}H_{0}\left(t-t_{0}\right).
\end{split} \label{solfora}
\end{equation}
As we did with the general solution, the dynamics of the universe will be studied by considering the Hubble parameter expressed by the Eq.(\ref{solforH}). The deceleration parameter can be written as
\begin{equation}
q(z)=-1+\frac{m_{1}C_{1}+m_{2}C_{2}\left(1+z\right)^{m_{2}-m_{1}}}{C_{1}+C_{2}\left(1+z\right)^{m_{2}-m_{1}}}. \label{qofz}
\end{equation}

We will study the behavior of both parameters at early times and at very far future. Considering the Eqs.(\ref{defofm1}) and (\ref{defofm2}), it follows that $m_{1}>0$ and $m_{1}>m_{2}$ hold. Therefore, at early times the Hubble and deceleration parameters behaves following the simple expressions
\begin{equation}
H\left(z\rightarrow\infty\right)\rightarrow
C_{1}H_{0}\left(1+z\right)^{m_{1}}, \label{Hearly}
\end{equation}
\begin{equation}
q\left(z\rightarrow\infty\right)\rightarrow -1+m_{1}, \label{qofzearly}
\end{equation}
while for very far future they behave as
\begin{equation}
H\left(z\rightarrow -1\right)\rightarrow C_{2}H_{0}\left(1+z\right)^{m_{2}}, \label{Hlate}
\end{equation}
\begin{equation}
q\left(z\rightarrow -1\right)\rightarrow -1+m_{2}. \label{qofzlate}
\end{equation}

\begin{figure}
\centering
\subfigure[(Color online) Allowed values that leads to a positive Hubble parameter, compatible with Eq.(\ref{constraintHpositive}).]{\includegraphics[scale=0.68]{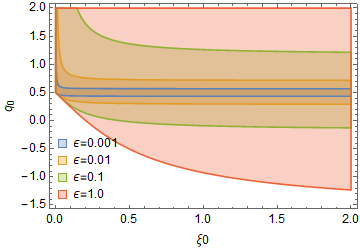}}
\vspace{0.05cm}
\subfigure[ (Color online) Allowed values that leads to a positive (negative) Hubble parameter at early (late) time, see Eq.(\ref{constraintC2negative}).]{\includegraphics[scale=0.68]{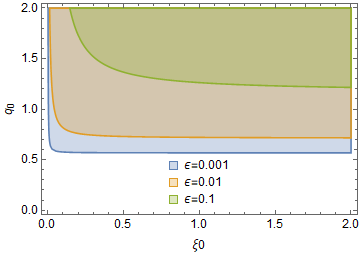}}
\vspace{0.05cm}
\subfigure[(Color online) Allowed values that leads to a negative (positive) Hubble parameter at early (late) time, see Eq.(\ref{constraintC1negative}).]{\includegraphics[scale=0.68]{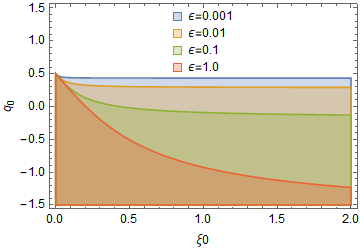}}
\caption{Comparative graphics of the allowed values of the parameters $q_{0}$ and $\xi_{0}$ for fixed $\epsilon$, which lead to a positive $H$ in the whole range of cosmological time(a), positive $H$ only at early times and negative $H$ at late times (b), and negative (positive) $H$ at early (late) time (c).}
\label{fig:constraintq0}
\end{figure}

In the latter case, the Hubble parameter is not necessarily positive during the cosmic evolution. In fact, from Eq.(\ref{solforH}) one sees that the Hubble parameter is zero for
\begin{equation}
(1+z)^{m_{2}-m_{1}}=-\frac{C_{1}}{C_{2}}. \label{Hzero}
\end{equation}
Because $1+z>0$, from Eqs.(\ref{Hearly}) and (\ref{Hlate}), it follows that the Hubble parameter will always be positive for $C_{1}>0$ and $C_{2}>0$, and always negative for $C_{1}<0$ and $C_{2}<0$. Positive at early times and negative at late times for $C_{1}>0$ and $C_{2}<0$, and negative at early times and positive at late times for $C_{1}<0$ and $C_{2}>0$. Note that $C_{1}>0$ requires the following constraint for the deceleration parameter (see Eq.(\ref{defofc1}))
\begin{equation}
q_{0}>\frac{1}{2}+\frac{\sqrt{3}}{2\xi_{0}}\left(\epsilon-\sqrt{6\xi_{0}^{2}\epsilon+\epsilon^{2}}\right), \label{constraintC1positive}
\end{equation}
and $C_{2}>0$ requires the following constraint for the deceleration parameter
\begin{equation}
q_{0}<\frac{1}{2}+\frac{\sqrt{3}}{2\xi_{0}}\left(\epsilon+\sqrt{6\xi_{0}^{2}\epsilon+\epsilon^{2}}\right). \label{constraintC2positive}
\end{equation}
Hence, a positive Hubble parameter requires a deceleration parameter bounded according to (see Fig.\ref{fig:constraintq0}(a))
\begin{equation}
\epsilon-\sqrt{6\xi_{0}^{2}\epsilon+\epsilon^{2}}<\frac{2\xi_{0}}{\sqrt{3}}\left(q_{0}-\frac{1}{2}\right)<\epsilon+\sqrt{6\xi_{0}^{2}\epsilon+\epsilon^{2}}, \label{constraintHpositive}
\end{equation}
while a negative Hubble parameter, in the whole region is clearly not possible. On the other hand, a positive Hubble parameter at early times and negative at late times requires that $q_{0}$ fulfills the condition of Eq.(\ref{constraintC1positive}) and
\begin{equation}
q_{0}>\frac{1}{2}+\frac{\sqrt{3}}{2\xi_{0}}\left(\epsilon+\sqrt{6\xi_{0}^{2}\epsilon+\epsilon^{2}}\right), \label{constraintC2negative}
\end{equation}
whose intersection is showed in Fig.\ref{fig:constraintq0}(b). Finally, a negative Hubble parameter at early times and positive at late times requires that $q_{0}$ fulfilled the condition
(\ref{constraintC2positive}) and
\begin{equation}
q_{0}<\frac{1}{2}+\frac{\sqrt{3}}{2\xi_{0}}\left(\epsilon-\sqrt{6\xi_{0}^{2}\epsilon+\epsilon^{2}}\right), \label{constraintC1negative}
\end{equation}
whose intersection is showed in see Fig.\ref{fig:constraintq0}(c).

The behavior of Eq.(\ref{Hearly}) depends on the exponent $m_{1}$ which we already mentioned that it is positive. Furthermore, this exponent has the following constraint
\begin{equation}
m_{1}\geq\frac{3}{2}. \label{exponentconstraintearly}
\end{equation}
Therefore, the Hubble parameter is positive at early times and decreases when the scale factor grows up to a positive non-zero value. If the condition (\ref{constraintC2negative}) is fulfilled, then the Hubble parameter decreases up to a negative value. On the other hand, if the condition (\ref{constraintC1negative}) is fulfilled, then the Hubble parameter is negative and increases when the scale factor grows up to a positive non-zero value. This behaviors correspond to a decelerated expansion, as can be see from Eqs.(\ref{qofzearly}) and (\ref{exponentconstraintearly}), which lead to a value of the deceleration parameter $q\geq 1/2$.

On the other hand, the behavior of Eqs.(\ref{Hlate}) and (\ref{qofzlate}) depends of the exponent $m_{2}$, which will be positive only if
\begin{equation}
3\left(1-2\epsilon\right)\xi_{0}+2\sqrt{3}\epsilon>0. \label{conditionpower}
\end{equation}
Therefore, if $0<\epsilon<1/2$, then $m_{2}>0$ and from
Eq.(\ref{Hlate}) we will have a Hubble parameter that at late times
continues decreasing, getting closer to zero and for the positives,
if we fulfilled with the conditions (\ref{constraintHpositive}) or
(\ref{constraintC1negative}). If we fulfilled the condition
(\ref{constraintC2negative}), then the Hubble parameter goes to zero
at late times but from negatives values. The same behavior for the
Hubble parameter at late time is possible when $1/2<\epsilon\leq 1$,
if and only if, $\xi_{0}$ is given by the constraint
\begin{equation}
\xi_{0}<\frac{2\epsilon}{\sqrt{3}\left(2\epsilon-1\right)}. \label{constraintxi0power}
\end{equation}
If $0<\epsilon<1/2$, then $m_{2}<0$ when the condition
(\ref{constraintxi0power}) is violated and from Eq.(\ref{Hlate}) we
will have a Hubble parameter that at late times tends to positive
infinite value, if fulfills the restrictions
(\ref{constraintHpositive}) or (\ref{constraintC1negative}), or
tends to a negative infinite value if the restriction of
Eq.(\ref{constraintC2negative}) is fulfilled. Finally, if
$0<\epsilon<1/2$, the special case $m_{2}=0$ arises, in the
particular case where the inequality (\ref{constraintxi0power})
becomes and equality and from Eq.(\ref{Hlate}) a constant Hubble
parameter at late times is obtained.

\begin{figure}[ht]
\centering
\subfigure[\hspace{0.1cm}Plot of the deceleration parameter as a function of the redshift. The solid line corresponds to $\epsilon=0.7$, $\xi_{0}=0.9$ and $q_{0}=-0.6$; the dashed line to $\epsilon=0.3$, $\xi_{0}=0.5$ and $q_{0}=0.1$; and finally, the dashed-dotted line to $\epsilon=0.1$, $\xi_{0}=0.2$ and $q_{0}=0.4$.]{\includegraphics[scale=0.68]{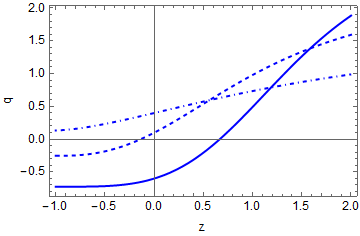}}
\vspace{0.05cm}
\subfigure[\hspace{0.1cm}Contour plot of the allowed values for the free parameters $\xi_{0}$ and $\epsilon$, which leads to a transition between an accelerated expansion to a decelerated one occurring at $z=0.64$ and for $q_{0}=-0.6$.]{\includegraphics[scale=0.68]{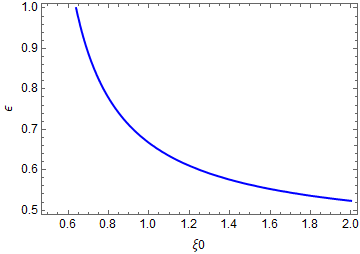}}
\caption{Plot of the deceleration parameter as a function of the redshift for fixed $\gamma=1.1$ (a), and contour plot of the allowed values for free parameters that lead to a transition between the decelerated expansion to an accelerated one (b).}
\label{fig:q0behaviorgamma1}
\end{figure}

From Eq.(\ref{qofzlate}) it follows this last behavior leads to a
stage of accelerated expansion when $m_{2}<1$, and this is only
possible under the condition
\begin{equation}
\xi_{0}>\frac{2\sqrt{3}\epsilon}{18\epsilon-1}, \label{constraintxi0m1}
\end{equation}
for
\begin{equation}
1/18<\epsilon\leq 1. \label{constraintxi0m1epsilon}
\end{equation}
If one of the above conditions is not fulfilled, then the behavior
of the Hubble parameter leads to a stage of decelerated expansion.
The transition between the accelerated expansion and the decelerated
one occurs at the redshift value
\begin{equation}
z_{q=0}=-1+\left[-\frac{C_{1}\left(1-m_{1}\right)}{C_{2}\left(1-m_{2}\right)}\right]^{1/(m_{2}-m_{1})}.
\end{equation}
In Fig.\ref{fig:q0behaviorgamma1}, the behavior of the
deceleration parameter $q (z)$ is displayed in terms of the free parameters 
$\xi_{0}$ and $\epsilon$ for a positive Hubble parameter.

\subsection{Thermodynamics properties of the solution}
For $\gamma=1$ Eq.(\ref{entropyproduction1}) takes the form
\begin{equation}
\dfrac{dS}{dt}=\frac{3H^{2}a^{3}}{T_{0}n_{0}}\left(2\dfrac{dH}{da}a+3H\right), \label{entropyproduction2}
\end{equation}
and using the Eq.(\ref{solforH}), this can be written as
\begin{equation}
\begin{split}
& \dfrac{dS}{dt}=\frac{3H_{0}H^{2}(1+z)^{-3}}{T_{0}n_{0}}\times \\
& \left[C_{1}\left(1+z\right)^{m_{1}}\left(3-2m_{1}\right)+C_{2}\left(1+z\right)^{m_{2}}\left(3-2m_{2}\right)\right].
\label{entropyofa}
\end{split}
\end{equation}
Due to the second law of Thermodynamics, the above derivative must
be a non-negative function of the cosmological time. As the first
factor in the above expression is positive, a non-negative entropy
production requires
\begin{equation}
\left[C_{1}(1+z)^{m_{1}}\left(3-2m_{1}\right)+C_{2}(1+z)^{m_{2}}\left(3-2m_{2}\right)\right]> 0. \label{entropycondition}
\end{equation}
As we have done for the Hubble parameter, we are going to analyzed
the above condition only for early and very far future times. This
is why we only considered the strict inequality in the
Eq.(\ref{entropycondition}). If $z\rightarrow\infty$ the above term
in bracket tends to the expression
\begin{equation}
C_{1}\left(3-2m_{1}\right)>0, \label{asymptotics1}
\end{equation}
but from Eqs.(\ref{defofm1}) and (\ref{defofm2}) it follows
\begin{equation}
3-2m_{1}=-\frac{\sqrt{3}}{2\xi_{0}}\left(\epsilon+\sqrt{6\xi_{0}^{2}\epsilon+\epsilon^{2}}\right)<0,
\label{asymptoticm1}
\end{equation}
\begin{equation}
3-2m_{2}=\frac{\sqrt{3}}{2\xi_{0}}\left(\sqrt{6\xi_{0}^{2}\epsilon+\epsilon^{2}}-\epsilon\right)>0,
\label{asymptoticm2}
\end{equation}
so, the Eq.(\ref{asymptoticm1}) shows that a positive entropy
production at early times requires a negative constant $C_{1}$,
which contradicts the content of the
Eq.(\ref{constraintC1positive}), i.e., a positive entropy production
at early times necessarily implies a negative Hubble parameter. On
the other hand, for $z\rightarrow -1$ the terms in brackets in
Eq.(\ref{entropycondition}) tends to the expression
\begin{equation}
C_{2}\left(3-2m_{2}\right)>0, \label{asymptotics2}
\end{equation}
and the Eq.(\ref{asymptoticm2}) shows that a positive entropy
production at early times requires a positive constant $C_{2}$, that
is the condition indicated in the Eq.(\ref{constraintC2positive}),
i.e., a positive entropy production at late times necessarily
requires a positive Hubble parameter at this times. Thus, a positive
entropy production for all the cosmic evolution is only possible for
a Hubble parameter that is negative at early times an positive at
late times. In the other hand, a positive Hubble parameter for all
the cosmic evolution leads to a negative entropy production at early
times and positive at late times. From Eq.(\ref{entropyofa}) it can
be see that the change of sign in the entropy production occurs at
redshift given by
\begin{equation}
z_{ds/dt=0}=-1+\left[-\frac{C_{1}\left(3-2m_{1}\right)}{C_{2}\left(3-2m_{2}\right)}\right]^{1/(m_{2}-m_{1})}. \label{zentropyequalcero}
\end{equation}
A numerical calculation of Eq.(\ref{zentropyequalcero}) indicates us
that the the transition from a negative entropy production to a
positive one may occurs at z in the range $1<z<5$ choosing values of
epsilon between $0.5$ and $0.7$, and $0.8<\xi_{0}<2$. The result is
similar to the case of $\gamma\neq 1$ but in this case, the value of
the redshift es lower than the value for the general case, for the
same values of $\epsilon$ and $\xi_{0}$; even so, the intervals are
the same. The conclusion is this case is similar to the former case
$\gamma\neq 1$.

\subsection{Special cases of the particular solution}
In the Eq.(\ref{solforH}) there are two particular cases:
i) $C_{1}=0$ and $C_{2}=1$ and ii) $C_{1}=1$ and $C_{2}=0$.
These particular cases were not addressed so far because they lead to quite different physical scenarios that we will discuss in this section.

From Eq.(\ref{defofc1}) we see that in the case i) the deceleration parameter has the particular value
\begin{equation}
q_{0}=m_{2}-1, \label{constraintC10}
\end{equation}
which correspond to Eq.(\ref{constraintC1positive}) but with
\begin{equation}
q_{0}=\frac{1}{2}+\frac{\sqrt{3}}{2\xi_{0}}\left(\epsilon-\sqrt{6\xi_{0}^{2}\epsilon+\epsilon^{2}} \right). \label{q0special}
\end{equation}
This leads to a Hubble parameter as a function of the scale factor of the form
\begin{equation}
H\left(a\right)=H_{0}a^{-m_{2}}, \label{HC10}
\end{equation}
which is always positive during cosmic evolution. The scale factor can be obtained straightforwardly, and is given by
\begin{equation}
H_{0}t+C=\int{a^{m_{2}-1}da}, \label{intHC10}
\end{equation}
where $C$ is an integration constant.

Taking $m_{2}=0$ in Eq.(\ref{HC10}) we obtain a de Sitter type
expansion with a constant Hubble parameter. The scale factor, with
the initial condition $a(t_{0})=1$, is given as a function of time
by the expression
\begin{equation}
a\left(t\right)=e^{H_{0}\left(t-t_{0}\right)} \label{aC10DeSitter}.
\end{equation}

For $m_{2}\neq 0$, the scale factor as a function of the cosmic time is given by
\begin{equation}
a\left(t\right)=\left[H_{0}\left(t-t_{0}\right)m_{2}+1\right]^{1/m_{2}}. \label{aC10}
\end{equation}
Inserting this expression into Eq.(\ref{HC10}), one obtains the following Hubble parameter
\begin{equation}
H\left(t\right)=\frac{H_{0}}{H_{0}\left(t-t_{0}\right)m_{2}+1}. \label{HC10t}
\end{equation}
In order to avoid nonphysical scale factors. The solution
(\ref{aC10}) represents an universe with an origin at time
$t=t_{0}-1/(H_{0}m_{2})$ and with an accelerated expansion for
$0<m_{2}<1$ and a decelerated expansion for $m_{2}>1$. The case
$m_{2}=1$ is clearly an universe with constant rate of expansion
during the whole cosmic evolution. Finally, for $m_{2}<0$ the
Eq.(\ref{aC10}) can be rewritten as
\begin{equation}
a\left(t\right)=\frac{1}{\left[1-H_{0}\left(t-t_{0}\right)|m_{2}|\right]^{1/|m_{2}|}}, \label{aemergent}
\end{equation}
where clearly one needs to impose
\begin{equation}
t<t_{0}+\frac{1}{H_{0}|m_{2}|}. \label{constrainttbigrip}
\end{equation}
In this case the Eq.(\ref{aemergent}) represent an emergent
universe with an accelerated expansion at late times and a Big Rip at the time $t_{BR}=t_{0}+1/(H_{0}|m_{2}|)$.

Let us see now the behavior of the above solution in terms of their entropy production. In the case i) Eq.(\ref{entropyofa}), for the
particular values $C_{1}=0$ and $C_{2}=1$, gives the entropy production as a function of the scale factor, which yields
\begin{equation}
\dfrac{dS}{dt}=\frac{3H_{0}^{3}}{T_{0}n_{0}}\left(3-2m_{2}\right)a^{3\left(1-m_{2}\right)}, \label{entropyC10}
\end{equation}
which indicates that the entropy production is always positive since
$3-2m_{2}>0$ by Eq.(\ref{asymptoticm2}). Within this range of the
parameter $m_{2}$ we have the cosmological scenarios with
accelerated expansion ($0<m_{2}<1$), with decelerated expansion
($m_{2}>1$) and expansion at constant rate ($m_{2}=1$). These
special cases have the particularity of the absent of transition
from a decelerated phase to an accelerated expansion. I this sense
they present a well behavior in terms of the thermodynamics but they
are unable to model the universe like the $\Lambda CDM$ model where
a transition from the DM dominated era to the DE era naturally
appears.

The other case, $C_{2}=0$ and hence $C_{1}=1$ will not be addressed explicitly as it drives to a cosmic evolution with nonphysical negative entropy production.

\section{Conclusions}

We first point out that in the general novel solution for arbitrary $\gamma$, the
entropy production is negative at early times while it is positive for late
times, being the Hubble parameter positive. Moreover, a
transition from a decelerated phase to an accelerated one is only
possible for large values of $\epsilon$ (see
Eqs.(\ref{constraintxi0acceleratedgamma}) and
(\ref{constraintepsilonacceleratedgamma})).

In the particular case $\gamma =1$, one sees that the Hubble
parameter can be positive or negative at early times as well as late
times, depending on the election of the deceleration parameter
$q_{0}$. It is worth mentioning that this accelerated expansion is
compatible with a positive Hubble parameter at late times, and
therefore one needs a $q_{0}$ value that fulfills the conditions
indicated above in Eq.(\ref{constraintC1negative}), which is
possible for some values of the free parameters. In particular, for
$1/18<\epsilon\leq 1$ an accelerated expansion arises if and only if, $\xi_{0}$
satisfies the inequality of Eq.(\ref{constraintxi0m1}). Nevertheless,
if $0<\epsilon\leq 1/18$, then the accelerated expansion will not be
possible, independent of the value of $\xi_{0}$. From
the thermodynamical point of view, we have found that the entropy
production for this model can be positive or negative depending on
the Hubble parameter at early and late times. As our model contains 
only one cold fluid as the main component of the universe, it should 
only be considered as an adequate
approximation for the late time evolution, where cold DM dominates.
In this sense, our model cannot expected to be fairly representative
of the early times evolution, where ultrarelativistic matter
dominates, which implies that the positiveness of the 
entropy production at late times must hold.

As a summary, the solution for $\gamma =1$ and $1/18<\epsilon\leq 1$
should be considered as a suited scenario of a cosmic evolution: it has a
transition between decelerated and accelerated expansions, and at
the same time, it has a positive entropy production at late times.
Its non-physical negative entropy production at early times should
not be considered a reason to discard it, as it was argued above. 

One unwanted feature of the solution is that the accelerated expansion 
at late times happens only for a relative large $\epsilon$ value, which
implies that the non-adiabatic contribution from dissipation to the speed of sound
would be to large.

\begin{acknowledgments}
This article was partially supported by Dicyt from Universidad de Santiago de Chile, through Grants $N^{\circ}$ $041831PA$ (G.P.) and $N^{\circ}$ $041831CM$ (N.C.). E.G. was supported by CONICYT-PCHA/Doctorado Nacional/2016-21160331.
\end{acknowledgments}

\end{document}